\documentclass[hyper]{JHEP} 
\usepackage{epsfig}
\usepackage{amsmath}
 \usepackage{slashed}
\newcommand\fverb{\setbox\pippobox=\hbox\bgroup\verb}
\newcommand\fverbdo{\egroup\medskip\noindent%
            \fbox{\unhbox\pippobox}\ }

\newcommand\fverbit{\egroup\item[\fbox{\unhbox\pippobox}]}
\newbox\pippobox
\title{$(m,n)-$ String in    $(p,q)$-String
and $(p,q)-$Five Brane Background}

\preprint{\hepth{}}

\author{
Josef Kluso\v{n} \\
Institute for Theoretical Physics  and Astrophysics\\
Faculty of Science, Masaryk University\\
Kotl\'{a}\v{r}sk\'{a} 2, 611 37, Brno\\
Czech Republic\\
E-mail: \email{klu@physics.muni.cz}}

\abstract{We study  dynamics of $(m,n)-$string in $(p,q)-$five brane
and $(p,q)-$string background. We determine world-volume stress
energy tensor and we analyze the dependence of the string's dynamics
on the values of the charges $(m,n)$ and the value of the angular
momentum.} \keywords{D-brane, Supergravity Background}
\def\det{\mathrm{det}}

\def\mP{\mathcal{P}}

\def\tr{\mathrm{Tr}}

\def\mF{\mathcal{F}}

\def\mM{\mathcal{M}}

\def\tkappa{\tilde{\kappa}}

\def\pb  #1{\left\{#1\right\}}

\def\bm{\mathbf{m}}
\def\bB{\mathbf{B}}
\newcommand{\bA}{{\bf A}}

\newcommand{\mH}{\mathcal{H}}
\newcommand{\mG}{\mathcal{G}}

\def\mK{\mathcal{K}}

\newcommand{\mL}{\mathcal{L}}

\newcommand{\bH}{\mathbf{H}}

\begin{document}

\section{Introduction and Summary}
Low energy effective actions of superstring theories have reach
spectrum of solutions that preserve some fractions of supersymmetry,
for review see for example
\cite{Duff:1994an,Smith:2002wn,Stelle:1996tz,Ortin:2004ms}. These
objects have property that they  are sources of various form fields
that are presented in supergravity theories. Further, fundamental
string, D-brane and NS5-brane solutions preserve one half of the
space-time supersymmetries and can be considered as the building
block of other solutions. For example, taking
 intersection of these
configurations  we get backgrounds that preserve some fractions of
supersymmetry \cite{Tseytlin:1996bh}. Another possibility is to
generate new solutions using U-duality symmetry of M-theory (For
review see for example \cite{Obers:1998fb}) which is basically the
symmetry of M-theory on its maximally supersymmetric toroidal
compactifications. For example, M-theory compactified on two torus
possesses $SL(2,Z)$ symmetry which turns out to be non-perturbative
$SL(2,Z)$ duality of type IIB theory.  More precisely, it is well
known that low effective action of type IIB supergravity written in
Einstein frame is invariant under $SL(2,R)$ duality. Special case of
$SL(2,R)$ transformation is S-duality transformation that roughly
speaking transforms theory at week coupling to the strong coupling.
The fact that the type IIB supergravity action is invariant under
this symmetry suggests the possibility how to generate new
supergravity solutions when we apply $SL(2,R)$ rotation on known
supergravity solutions, as for example fundamental string or
NS5-brane backgrounds. Such a procedure was firstly used in a famous
paper \cite{Schwarz:1995dk} where the manifestly $SL(2,R)$ covariant
supergravity solution corresponding to $(p,q)$-string was found. The
extension of this analysis to the case of NS5-brane was performed in
\cite{Lu:1998vh} when $SL(2,Z)$ covariant expression for
supergravity solutions corresponding to $(p,q)-$five brane was
derived \footnote{For non-extremal form of these solutions, see
\cite{Bueno:2014zja}.} . These backgrounds are very interesting and
certainly deserve to be studied further. In particular, it is well
known that the continuous classical symmetry group $SL(2,R)$ of type
IIB supergravity cannot be a symmetry of the full string theory when
non-perturbative effects break it to a discrete subgroup $SL(2,Z)$.
To see this more clearly  note that fundamental string carries one
unit of NSNS two form charge and hence this charge has to be
quantized in integer units. On the other hand $SL(2,R)$
transformations maps a fundamental string into a string with
$d-$units of this charge where $d$ is an entry of $SL(2,R)$ matrix.
From this result we conclude that $d-$has to be integer. In the
similar way we can argue that $SL(2,R)$ symmetry of the low energy
effective action has to be broken to its $SL(2,Z)$ subgroup when
fundamental string  is mapped under this duality to $(p,q)-$string
that carries $p$ charge of NSNS-two form and $q$ charge of Ramond
Ramond two form \cite{Tseytlin:1996it}. It was also shown in
\cite{Tseytlin:1996it}
 that the type IIB  string effective action together with $(p,q)-$string
action is covariant under $SL(2,R)$ transformations. However
 the fact that $(p,q)$ string has to map to another $(p',q')-$
string where $p',q'$ are integers suggests that the full symmetry
group of the combined action breaks to $SL(2,Z)$. On the other hand
solutions found in \cite{Schwarz:1995dk,Lu:1998vh} were determined
using the $SL(2,R)$ matrices so that it is interesting to analyze
the problem of $(m,n)-$string probe in such a background and this is
precisely the aim of this paper.

We begin with the D1-brane action that we rewrite into a manifestly
covariant $SL(2,Z)$ form, for related analysis see
\cite{Lozano:1997cy} and for very elegant formulation of manifestly
$SL(2,Z)$ covariant superstring, see
\cite{Cederwall:1997ts,Townsend:1997kr}. Now using the fact that
$(p,q)-$five and fundamental string solutions were derived using
$SL(2,R)$ transformations we can map the problem of the dynamics of
$(m,n)$-string in this background to the problem of the analysis of
$(m',n')-$ string in the original NS5-brane and fundamental string
background with crucial exception that the harmonic functions that
define these solutions have constant factors that differ from the
factors that define NS5-brane and fundamental string solutions. It
is also important to stress that now $(m',n')$ are not integers but
depend on $p,q$ and also on asymptotic values of dilaton and
Ramond-Ramond zero form. We mean that this is not quite satisfactory
resort and one can ask the questions whether it would be possible to
find $(p,q)$-string and five brane backgrounds that are derived from
the NS5-brane and fundamental string background through manifest
$SL(2,Z)$ transformations when in probe $(m,n)-$string will
transform in an appropriate way.  This problem is currently under
study and we return to it in near future. We rather focus on the
dynamics of the probe $(m,n)$-string in the backgrounds
\cite{Schwarz:1995dk,Lu:1998vh}, following very nice analysis
introduced in \cite{Kutasov:2004dj}. Using manifest $SL(2,Z)$
covariant formulation of a probe $(m,n)$-string we can analyze the
time evolution of homogeneous time dependent string in given
background. We determine components of the world-sheet stress energy
tensor and study its time evolution. The properties of this stress
energy tensor and the dynamics of the probe depends on the values of
$m,n$ and hence our results can be considered as the generalization
of the analysis performed in \cite{Kutasov:2004dj}.

As the next step we analyze the dynamics of the probe $(m,n)-$
string in the background of $(p,q)-$macroscopic string. Thanks to
the form of the solution \cite{Schwarz:1995dk} we formulate this
problem as the analysis of the dynamics of $(m',n')-$string in the
background of fundamental string. This problem was studied
previously in \cite{Bak:2004tp} but we focus on different aspect of
the dynamics of the probe. Explicitly we will be interested in the
behavior of the probe where the difference between its energy and
the rest energy is small. We find that the potential is flat which
is in agreement with the fact that the string probe in the
fundamental string background can form marginal bound state with the
strings that are sources of this background. We also analyze the
situation with non-zero angular momentum and we find that there is a
potential barrier that does not allow the probe string   to move
towards to the horizon. These results are in agreement with the
analysis performed in
 \cite{Bak:2004tp}.

The organization of this paper is as follows. In the next section
(\ref{second}) we review $SL(2,R)$ duality of type IIB low energy
effective action. We also introduce manifestly $SL(2,R)$ covariant
action for $(m,n)-$string. In section (\ref{third}) we study the
dynamics of this string in the background of $(p,q)-$five brane.
Finally in section (\ref{fourth}) we study dynamics of
$(m,n)-$string in the background of $(p,q)-$string.

\section{$SL(2,R)$-Covariance of type IIB Low Energy Effective Action} \label{second}
 The type IIB theory has two three-form field
strengths $H=dB, F=dC^{(2)}$, where  $H$ corresponds to NSNS three
form while $F$ belongs to RR sector and does not couple to the usual
string world-sheet. Type IIB theory has also two scalar fields that
can be combined into a complex field $\tau= \chi+ie^{-\Phi}$. The
dilaton $\Phi$ is in the NSNS sector while $\chi$ belongs to the RR
sector. The other bose fields are the metric $g_{\mu\nu}$ and
self-dual five form field strength $F_5$ that we set zero in  this
paper.  Then it is possible to write down a covariant form of the
bosonic part of type IIB effective action
\begin{equation}\label{SIIB}
S_{IIB}=\frac{1}{2\tkappa^2_{10}}\int d^{10}x
\sqrt{-g}(R+\frac{1}{4} \tr (\partial_\mu \mM
\partial^\mu \mM^{-1})-\frac{1}{12}\bH^T_{\mu\nu\sigma} \mM
\bH^{\mu\nu\sigma}) \ ,
\end{equation}
where $\tkappa^2_{10}=\frac{1}{4\pi}(4\pi^2\alpha')^4$ and
 where we
have combined $B,C^{(2)}$ into
\begin{equation}
\bH=d\bB=\left(\begin{array}{cc} dB \\
dC^{(2)} \\ \end{array}\right) \ ,
\end{equation}
and where
\begin{eqnarray}
\mM&=&e^{\Phi}\left(\begin{array}{cc} \tau \tau^* & \chi \\
\chi & 1 \\ \end{array}\right)= e^{\Phi}\left(\begin{array}{cc}
\chi^2+e^{-2\Phi} & \chi \\
\chi & 1 \\ \end{array}\right) \ . \quad
\nonumber \\
\end{eqnarray}
The  action (\ref{SIIB}) has manifest invariance under the global
$SL(2,R)$ transformation
\begin{equation}\label{deftrB}
\hat{\mM}=\Lambda \mM \Lambda^T \ ,  \hat{\bB}=(\Lambda^T)^{-1} \bB
\ ,
\end{equation}
where
\begin{equation}
\Lambda=\left(\begin{array}{cc} a & b \\
c& d \\ \end{array}\right) \ .
\end{equation}
It is well known that all string theories contain fundamental string
and magnetic dual NS5-brane as solutions of the equations of motion
of its low energy effective actions. Using a manifest $SL(2,R)$
covariance of type IIB effective action it is possible to derive
solutions corresponding to $(p,q)$-five brane \cite{Lu:1998vh} and
fundamental string \cite{Schwarz:1995dk}. It will be certainly
interesting to analyze properties of given background with the help
of the appropriate probe which will be probe $(m,n)-$string. For
that reason  we introduce a manifestly covariant form of
$(m,n)-$string action.
\subsection{$(m,n)-$String Action}
In this section we formulate the action for the $(m,n)-$ string.
Even if such a formulation is well known
\cite{Tseytlin:1996it,Lozano:1997cy,Cederwall:1997ts,Townsend:1997kr,Bergshoeff:2006gs}
we derive this action  in a slightly different way with the help of
the  Hamiltonian formalism which will be also useful for the
analysis of the dynamics of probe $(m,n)-$string in $(p,q)-$five and
$(p,q)-$string background.

To begin with we introduce an action for $n$ coincident D1-branes in
general background
\begin{eqnarray}\label{D1branegen}
S&=&-nT_{D1}\int d\tau d\sigma e^{-\Phi}\sqrt{-\det
\bA}+\nonumber \\
&+&nT_{D1}\int d\tau d\sigma
((b_{\tau\sigma}+2\pi\alpha'\mF_{\tau\sigma})\chi+ c_{\tau\sigma}) \
,
\nonumber \\
\bA_{\alpha\beta}&=&G_{MN}\partial_\alpha x^M
\partial_\beta x^N+2\pi\alpha'\mF_{\alpha\beta}+B_{MN}
\partial_\alpha x^M\partial_\beta x^N \ , \nonumber \\
\mF_{\alpha\beta}&=&\partial_\alpha A_\beta-\partial_\beta A_\alpha
\
, \nonumber \\
\end{eqnarray}
where $x^M,M,N=0,1,\dots,9$ are embedding coordinates of D1-brane in
the background that is specified by the metric $G_{MN}$ and NSNS two
form  $B_{MN}=-B_{NM}$ together with Ramond-Ramond two form
$C^{(2)}_{MN}=-C^{(2)}_{NM}$. Note that we use capital letter
$G_{MN}$ for the string frame metric while $g_{MN}$ corresponds to
the Einstein frame metric.  We further consider background with
non-trivial dilaton $\Phi$ and RR zero form $\chi$. Further,
$\sigma^\alpha=(\tau,\sigma)$ are world-sheet coordinates  and
$b_{\tau\sigma},c_{\tau\sigma}$ are pull-backs of $B_{MN}$ and
$C_{MN}$ to the world-volume of D1-brane. Explicitly,
\begin{equation}
b_{\alpha\beta}\equiv B_{MN}\partial_\alpha x^M\partial_\beta x^N \
, \quad  c_{\tau\sigma}=C^{(2)}_{MN}\partial_\tau x^M\partial_\sigma
x^N  \ .
\end{equation}
Finally $T_{D1}=\frac{1}{2\pi\alpha'}$ is D1-brane tension and
$A_{\alpha},\alpha=\tau,\sigma$ is two dimensional gauge field that
propagates on the world-sheet of D1-brane.

It is useful to rewrite the action (\ref{D1branegen}) into the form
\begin{eqnarray}\label{D1actional}
S&=&-nT_{D1}\int d\tau d\sigma e^{-\Phi} \sqrt{-\det g-
(2\pi\alpha'\mF_{\tau\sigma}+b_{\tau\sigma})^2}+ \nonumber \\
&+&nT_{D1}\int d\tau d\sigma
((b_{\tau\sigma}+2\pi\alpha'\mF_{\tau\sigma})\chi+ c_{\tau\sigma})
\ , \nonumber \\
\end{eqnarray}
where $g_{\alpha\beta}=G_{MN}\partial_\alpha x^M
\partial_\beta x^N, \det g=g_{\tau\tau}g_{\sigma\sigma}-
(g_{\tau\sigma})^2$.
Now we  proceed  to the Hamiltonian formulation of the theory
defined by the action (\ref{D1actional}). First of all we derive
conjugate momenta to $x^M$ and $A_\alpha$ from (\ref{D1actional})
\begin{eqnarray}\label{defpM}
& &p_M=\frac{\delta L}{\delta \partial_\tau x^M}= nT_{D1}
\frac{e^{-\Phi}}{\sqrt{-\det g -(2\pi\alpha'F_{\tau\sigma}+
b_{\tau\sigma})^2}}(G_{MN}\partial_\alpha x^N g^{\alpha \tau}\det g+\nonumber \\
& &+(2\pi\alpha'F_{\tau\sigma}+ b_{\tau\sigma})B_{MN}\partial_\sigma
x^N)+nT_{D1}(\chi B_{MN}\partial_\sigma x^N+C^{(2)}_{MN}\partial_\sigma x^N) \ , \nonumber \\
& &\pi^\sigma=\frac{\delta L}{\delta
\partial_\tau A_\sigma}=\frac{ne^{-\Phi}(2\pi\alpha'F_{\tau\sigma}+
b_{\tau\sigma})}{\sqrt{-\det g -(2\pi\alpha'F_{\tau\sigma}+
b_{\tau\sigma})^2}}+n\chi\ , \quad \pi^\tau=\frac{\delta L}{\delta
\partial_\tau A_
\tau}\approx 0 \nonumber \\
\end{eqnarray}
and hence
\begin{eqnarray}
\Pi_M&\equiv&
p_M-\frac{\pi^\sigma}{(2\pi\alpha')}B_{MN}\partial_\sigma x^N
-nT_{D1}C^{(2)}_{MN}\partial_\sigma
x^N)= \nonumber \\
&=&nT_{D1} \frac{e^{-\Phi}}{\sqrt{-\det g
-(2\pi\alpha'F_{\tau\sigma}+
b_{\tau\sigma})^2}}G_{MN}\partial_\alpha x^N g^{\alpha \tau}\det g \
.
\nonumber \\
\end{eqnarray}
Using these relations it is easy to see that the bare Hamiltonian is
equal to
\begin{eqnarray}
H_B=\int d\sigma(p_M\partial_\tau x^M+\pi^\sigma \partial_\tau
A_\sigma-\mL)= \int d\sigma \pi^\sigma\partial_\sigma A_\tau
\nonumber \\
\end{eqnarray}
while we have three primary constraints
\begin{eqnarray}
& &\pi^\tau\approx 0 \ , \quad \mH_\sigma\equiv p_M\partial_\sigma x^M\approx 0 \ , \nonumber \\
& &\mH_\tau\equiv\frac{1}{T_{D1}} \Pi_M G^{MN}\Pi_N+
T_{D1}\left(n^2e^{-2\Phi}+\left(\pi^\sigma -n\chi\right)^2
\right)g_{\sigma\sigma}\approx 0
\ . \nonumber \\
\end{eqnarray}
  Including
these primary constraints to the definition of the Hamiltonian we
obtain an extended Hamiltonian in the form
\begin{equation}
H=\int d\sigma (\lambda_\tau\mH_\tau+\lambda_\sigma
\mH_\sigma-A_\tau\partial_\sigma\pi^\sigma+v_\tau \pi^\tau) \ ,
\end{equation}
where $\lambda_\tau,\lambda_\sigma,v_\tau$ are Lagrange multipliers
corresponding to the primary constraints $\mH_\tau\approx 0 \
,\mH_\sigma\approx 0 \ , \pi^\tau\approx 0$.
Now we have to check the stability of all constraints. The
requirement of the preservation of the primary constraint
$\pi^\tau\approx 0$ implies the secondary constraint
\begin{equation}
\mG=\partial_\sigma \pi^\sigma\approx 0 \ .
\end{equation}
In case of  the constraints $\mH_\tau,\mH_\sigma$ we can easily show
in the same way as in \cite{Kluson:2014uaa} that the constraints
$\mH_\tau,\mH_\sigma$ are first class constraints and hence they are
preserved during the time evolution.

An  action for $(m,n)-$string is derived  when we fix the gauge
generated by $\mG$ with the gauge fixing function
$A_\sigma=\mathrm{const}$. Then  the fixing of the gauge implies
that $\pi^\sigma=f(\tau)$ but the equation of motion for
$\pi^\sigma$ implies that $\partial_\tau \pi^\sigma=0$ and hence
$\pi^\sigma=m$, where $m$ is integer that counts the number of
fundamental string bound to $n$ D1-branes. After this partial gauge
fixing  the Hamiltonian density has the form
\begin{equation}\label{Hamn}
\mH_{(m,n)}=\int d\sigma (\lambda_\tau\mH_\tau+\lambda_\sigma
\mH_\sigma) \ .
\end{equation}
In order to find $(m,n)$-string action we derive Lagrangian density
corresponding to the Hamiltonian (\ref{Hamn}). Explicitly, from
 (\ref{Hamn}) we obtain   equations of motion for $x^M$
\begin{equation}
\partial_\tau x^M=\pb{x^M,H_{(m,n)}}=
2\lambda_\tau \frac{1}{ T_{D1}}G^{MN}\Pi_N+\lambda_\sigma
\partial_\sigma x^M
\end{equation}
and hence
\begin{eqnarray}\label{lagden}
& &\mL_{(m,n)}=p_M\partial_\tau x^M-\mH_{(m,n)}=\nonumber \\
&=&\frac{1}{2\pi\alpha'}\left(\frac{1}{4\lambda_\tau}(g_{\tau\tau}-2\lambda_\sigma
g_{\tau\sigma}+\lambda_\sigma^2 g_{\sigma\sigma}) -\lambda_\tau
(n^2e^{-2\Phi}+(m-n\chi)^2)g_{\sigma\sigma} +m b_{\tau\sigma}+n
c_{\tau\sigma}\right)\ . \nonumber \\
\end{eqnarray}
As the final step  we solve the equation of motion for
$\lambda_\tau$ and $\lambda_\sigma$ that follow from (\ref{lagden})
and we obtain
\begin{eqnarray}
\lambda_\sigma=\frac{g_{\tau\sigma}}{g_{\sigma\sigma}} \ , \quad
\lambda_\tau=\frac{1}{2g_{\sigma\sigma}\sqrt{n^2e^{-2\Phi}+(m-n\chi)^2}}\sqrt{-\det
g} \ . \nonumber \\
\end{eqnarray}
Inserting this result into the Lagrangian density
(\ref{lagden}) we obtain the action in manifestly
covariant  $SL(2,R)$ form
\begin{eqnarray}\label{actcovSL}
S&=&-T_{D1}\int d\tau d\sigma(\sqrt{\bm^T\mM^{-1} \bm }\sqrt{-\det
g_{MN}\partial_\alpha x^M\partial_\beta x^N}\nonumber \\
 &+&T_{D1}\int d\tau
d\sigma \bm^T \bB_{MN}\partial_\tau x^M\partial_\sigma x^N \ ,
\nonumber \\
\end{eqnarray}
where
\begin{equation}
\bm=\left(\begin{array}{cc} m \\
n \\ \end{array}\right) \ , \bB_{MN}=\left(\begin{array}{cc}
B_{MN} \\
C_{MN}^{(2)} \\
 \end{array}\right) \ ,
\end{equation}
and where $g_{MN}=e^{-\Phi/2}G_{MN}$ is Einstein-frame
metric. Since $\hat{\bB}=(\Lambda^T)^{-1}\bB$ we obtain that $\bm$
transforms as
\begin{equation}\label{transbm}
\hat{\bm}=\Lambda \bm \
\end{equation}
in order  the action (\ref{actcovSL}) to be  manifestly $SL(2,R)$
covariant.  On the other hand since $m,n$ count the number of
fundamental strings and D1-branes and hence have to be integers we
find that the non-perturbative duality group of type IIB superstring
theory is $SL(2,Z)$ which will have an important consequence for the
analysis of the  dynamics of $(m,n)-$string in $(p,q)-$five brane
and $(p,q)-$fundamental string background.
\section{$(m,n)-$String in the Background of $(p,q)$-Five
Brane}\label{third} We would like to analyze the dynamics of
$(m,n)-$string in the background of $(p,q)$-five brane that has the
form  \cite{Lu:1998vh}
\begin{eqnarray}\label{solRoy}
ds^2_E&=&(1+\frac{Q_{(p,q)}}{r^2})^{-1/4} \eta_{\mu\nu}dx^\mu
dx^\nu+
(1+\frac{Q_{(p,q)}}{r^2})^{3/4}dx^m dx^m \ , \nonumber \\
\lambda&=&\chi+ie^{-\Phi}=\frac{\chi_0 \triangle_{(p,q)}A_{(p,q)}+pq
e^{-\Phi_0} (A_{(p,q)}-1)+i\triangle_{(p,q)}
A^{1/2}_{(p,q)}e^{-\Phi_0} }{ p^2
e^{-\Phi_0}+A_{(p,q)}e^{\Phi_0}(\chi_0 p+q)^2} \ , \nonumber \\
H&=&dB=2p (2\pi\alpha')^2\epsilon_3 \ , \quad  F=dC_2=2q
(2\pi\alpha')^2\epsilon_3 \ ,
\nonumber \\
\end{eqnarray}
where
\begin{equation}\label{Qpqfive}
Q_{(p,q)}=\sqrt{\triangle_{(p,q)}}2\pi\alpha'=\sqrt{e^{-\Phi_0}p^2+(q+p\chi_0)^2
e^{\Phi_0}}2\pi\alpha' \ ,
\end{equation}
and where $\epsilon_3$ is volume form of the three sphere when we
express the line element of the transverse space $dx_m dx^m$ as
$dx_m dx^m=dr^2+r^2 d\Omega_3$. Note also that $x^\mu,\mu=0,\dots,5$
label directions along the world-volume of $(p,q)-$five brane.
Further $A_{(p,q)}$ is defined as
 \begin{equation}
 A_{(p,q)}=\left(1+\frac{Q_{(p,q)}}{r^2}\right)^{-1} \ ,
 \end{equation}
 and $ds^2_E$ means that this line element is expressed in Einstein
 frame metric.
Let us now consider probe $(m,n)-$string action
(\ref{actcovSL}) in given background.  The analysis of this
problem simplifies considerably when we realize how the solution
(\ref{solRoy}) was determined. Following
\cite{Lu:1998vh} and \cite{Schwarz:1995dk} we introduce $SL(2,R)$ matrix
\begin{eqnarray}
\Lambda=\triangle_{(p,q)}^{-1/2} \left(\begin{array}{cc}
e^{-\Phi_0}p+\chi_0 e^{\Phi_0}(q+p\chi_0) & -(q+p\chi_0)+ \chi_0
 p \\
e^{\Phi_0}(q+p\chi_0) &
 p \\
\end{array}\right) \ , \nonumber \\
\end{eqnarray}
where
\begin{equation}
\triangle_{(p,q)}=e^{-\Phi_0}p^2+(q+p\chi_0)^2 e^{\Phi_0} \ ,
\end{equation}
and where $\chi_0$ and $\Phi_0$ are asymptotic values of fields
$\Phi$ and $\chi$. Note that  the inverse matrix has the form
\begin{equation}
\Lambda^{-1}=\triangle^{-1/2}_{(p,q)} \left(\begin{array}{cc} p & q
\\
-e^{\Phi_0}(q+p\chi_0) & e^{-\Phi_0}p+\chi_0 e^{\Phi_0}(q+p\chi_0)
\\ \end{array}\right) \ . \nonumber \\
\end{equation}
Now with the help of this matrix we can write $\mM$ as
\cite{Lu:1998vh}
\begin{equation}
\mM=\Lambda(p,q)\left(\begin{array}{cc} \sqrt{A_{(p,q)}} & 0 \\
0 & \frac{1}{\sqrt{A_{(p,q)}}} \\ \end{array}\right)\Lambda^T(p,q)
\end{equation}
so that
\begin{equation}
\bm^T \mM^{-1} \bm=
 m'^2 \frac{1}{\sqrt{A_{(p,q)}}}+n'^2 \sqrt{A_{(p,q)}}
 \ ,  \nonumber \\
\end{equation}
where  \begin{eqnarray}\label{bmbar}
\bm'=\left(\begin{array}{cc} m'\\
n' \\
\end{array}\right)=\Lambda^{-1}(p,q)\bm=\triangle^{-1/2}_{(p,q)}
\left(\begin{array}{cc} pm+qn \\
e^{\Phi_0}(q+p\chi_0)(- m+n\chi_0)+ e^{-\Phi_0}pn
\\ \end{array}\right) \ .
\end{eqnarray}
It is interesting that   for the special values of $m,n$ equal to
\begin{equation}
m=-q \ , n=p
\end{equation}
we obtain
\begin{equation}
\bm'=\triangle^{-1/2}\left(\begin{array}{cc} 0 \\
e^{\Phi_0}(q+p\chi_0)^2+ e^{-\Phi_0}p^2 \\
\end{array}\right)=\left(\begin{array}{cc}
0 \\
\triangle^{1/2}_{(p,q)} \\ \end{array}\right) \ .
\end{equation}
Since $m'=0$ we can interpret this configuration as a
 pure D1-brane which  however does not have integer
charge. We also see from (\ref{bmbar}) that in order to find
configuration with $n'=0$  we have to demand that
 $\Phi_0=0=\chi_0$ and set
$m=p,n=q$
\begin{equation}
\bm'=\triangle^{-1/2}\left(\begin{array}{cc}p^2+q^2 \\
0 & \\ \end{array}\right)=
\left(\begin{array}{cc}
\sqrt{p^2+q^2} \\
0 \\ \end{array}\right) \ .
\end{equation}
%
%
%
Generally we see that the action for the probe $(m,n)-$string in
$(p,q)-$five brane background is equivalent to the action of
$(m',n')-$string in NS5-brane background with an important exception
that the harmonic function has the factor $Q_{(p,q)}$
(\ref{Qpqfive}) instead of the standard one that corresponds to the
number of NS5-branes. Note also that $m',n'$ depend on $m,n,p,q$ and
moduli $\Phi_0$ and $\chi_0$ as follows from (\ref{bmbar}).

 Let us now return to  the analysis of dynamics of
probe $(m,n)-$string in this background. It is convenient to impose
the static gauge
\begin{equation} x^0=\tau , \quad
x^1=\sigma \
\end{equation}
and introduce spherical coordinates in the transverse space
$\mathbf{R}^4$
\begin{equation}
x^1=r\cos \psi \ , x_1=r\sin\psi\cos\theta \ ,
x^3=r\sin\psi\sin\theta\cos\phi \ , x^4=r\sin\psi \sin
\theta\sin\phi
\end{equation}
so that volume element of $\Omega_3$ is equal to
\begin{equation}
d\Omega_3=\sin^2\psi \sin \theta d\psi \wedge d\theta \wedge  d\phi
\ .
\end{equation}
Using these equations we obtain that we have following components of
RR and NSNS two forms
\begin{eqnarray}
 B_{\psi\phi}&=&2p(2
\pi\alpha')^2\sin^2\phi \cos\theta \ , \quad
 C^{(2)}_{\psi\phi}=2q(2\pi\alpha')^2 \sin^2\psi \cos\theta \ .
\nonumber \\
\end{eqnarray}
Now we would like to derive the components of the stress energy
tensor $T_{\alpha\beta}$ for the gauge fixed theory. To do this we
temporary replace fixed two dimensional metric $\eta_{\alpha\beta}$
with two dimensional metric $\gamma_{\alpha\beta}$ and write the
gauge fixed action in the form
\begin{equation}
S_{fixed}=-T_{D1}\int d\tau d\sigma(\sqrt{\bm^T\mM^{-1} \bm
}A_{(p,q)}^{1/4}\sqrt{-\det \bA_{\alpha\beta}} +S_{WZ} \ ,
\end{equation}
where
\begin{equation}
\bA_{\alpha\beta}=\gamma_{\alpha\beta}+\frac{1}{A_{(p,q)}}\delta_{mn}
\partial_\alpha x^m\partial_\beta x^n+\delta_{
\alpha\beta}\partial_\alpha x^\alpha \partial_\beta x^\beta \ ,
\end{equation}
where $x^\alpha,\alpha=1,\dots,5$ label coordinates along the
world-volume of $(p,q)-$five brane. Then we define components of two
dimensional stress energy tensor as
\begin{equation}
T_{\alpha\beta}=-\frac{2}{\sqrt{-\det \gamma}}\frac{\delta
S_{fixed}}{\delta \gamma^{\alpha\beta}}=-\frac{T_{D1}}{\sqrt{-\det
\gamma}}\gamma_{\alpha\gamma}(\bA^{-1})^{\gamma\delta}\gamma_{\delta\beta}
\sqrt{-\det\bA}\sqrt{\bm^T\mM^{-1} \bm }A_{(p,q)}^{1/4} \ .
\end{equation}
Now we return back to the flat metric
$\gamma_{\alpha\beta}\rightarrow \eta_{\alpha\beta}$ and consider
pure time-dependent ansatz. As a result we obtain following
components of the world-sheet stress energy tensor
\begin{eqnarray}\label{TNS}
T_{\tau\tau}&=&
\frac{T_{D1} \sqrt{\bm^T\mM^{-1} \bm
}A_{(p,q)}^{1/4}}{\sqrt{1-\frac{1}{A_{(p,q)}}\partial_\tau x^m
\partial_\tau x_m-\partial_\tau x^\alpha\partial_\tau x_\alpha}} \ ,  \quad
T_{\tau\sigma}=0 \ , \nonumber \\
T_{\sigma\sigma}&=&-T_{D1} \sqrt{\bm^T\mM^{-1} \bm
}A_{(p,q)}^{1/4}\sqrt{1-\frac{1}{A_{(p,q)}}\partial_\tau x^m
\partial_\tau x_m-\partial_\tau x^\alpha \partial_\tau x_\alpha}\  \nonumber \\
\end{eqnarray}
that are generalization of the components of the stress energy
tensor of Dp-brane moving in NS5-brane background as were found in
\cite{Kutasov:2004dj}.
\subsection{Gauge Fixing in Hamiltonian formalism}
Now we proceed to the analysis of dynamics of
 probe $(m,n)-$string in  $(p,q)-$five brane background. It
turns out that it is useful to perform this analysis in the
canonical approach when we impose the static gauge using two  gauge
fixing functions
\begin{equation}
\mG_\tau=x^0-\tau \approx 0 \ , \quad \mG_\sigma=x^1-\sigma \approx
0 \ .
\end{equation}
These constraints have non-zero Poisson brackets with
$\mH_\tau\approx 0 \ , \mH_\sigma \approx 0$ so that they  are  the
second class constraints. As a result $\mH_\tau,\mH_\sigma$ vanish
strongly and can be solved for $p_0$ and $p_1$ respectively, where
we can relate $-p_0$ with the Hamiltonian density of gauge fixed
theory $\mH_{fix}$. To see this note that the action has the form
\begin{equation}
S=\int d\tau d\sigma (p_M\partial_\tau x^M-\mH)= \int d\tau d\sigma
(p_i\partial_\tau x^i+p_0)=\int d\tau d\sigma (p_i\partial_\tau
x^i-\mH_{fix}) \ .
\end{equation}
Now from $\mH_\sigma=0$ we obtain $p_1=-(p_i\partial_\sigma x^i)$
and from $\mH_\tau$ we find
\begin{eqnarray}\label{Hfix}
\mH_{fix}&=&\sqrt{-g_{00}\left(\Pi_1 g^{11}\Pi_1+\Pi_i g^{ij}\Pi_j
+T_{D1}^2 (m'^2e^{\Phi}+n'^2 e^{-\Phi})(g_{11}+g_{ij}
\partial_\sigma x^i\partial_\sigma x^j)\right)}\nonumber \\
&-&\frac{1}{2\pi\alpha'}
\bm^T \bB_{0M}\partial_\sigma x^M \ ,  \nonumber \\
\end{eqnarray}
where $i,j=2,\dots,9$.  The analysis simplifies further when we
presume that the embedding modes depend on $\tau$ only so that  the
Hamiltonian density (\ref{Hfix}) reduces into
\begin{eqnarray}
\mH^2_{fix}&=&A_{(p,q)}^{1/4}\left(p^\alpha p_\alpha
A_{(p,q)}^{-1/4}+A_{(p,q)}^{3/4}(p_r^2+ \frac{1}{r^2}p_\psi^2+
\frac{1}{r^2\sin^2\psi}p_\theta^2+\frac{1}{r^2
\sin^2\psi\sin^2\theta }p_\phi^2)+\right.\nonumber \\
& &\left.+T_{D1}^2(m'^2+ n'^2
A_{(p,q)})A^{-1/4}_{(p,q)}\right)\equiv \mK \ ,
\nonumber \\
\end{eqnarray}
where $p_\alpha,\alpha=2,3,4,5$ denote momenta along the
world-volume of $(p,q)-$five branes. Since they are conserved we
restrict ourselves to the case when $p_\alpha=0$. At the same time
we find that  $p_\phi$ is conserved as well and we denote this
constant as $p_\phi=L$.
On the other hand the equations of motion for $\theta,p_\theta$ have
the form
\begin{eqnarray}
\dot{\theta}&=&\pb{\theta,H_{fix}}=\frac{A_{(p,q)}p_\theta}{r^2\sin^2\psi
\sqrt{\mK}} \ , \nonumber \\
\dot{p}_\theta&=&\pb{p_\theta,H_{fix}}=\frac{A_{(p,q)}\sin\theta\cos\theta}
{r^2\sin^2\psi \sin^3\theta\sqrt{\mK}}p_\psi^2 \ . \nonumber \\
\end{eqnarray}
We see that this equation has the solution when
$\theta=\frac{\pi}{2}$ and $p_\theta=0$.  In the same way we find
that $p_\psi=0,\psi=\frac{\pi}{2}$ solve the equations of motion.
Finally we proceed to the analysis of the time evolution of $r$. The
equation of motion for $r$ gives
\begin{equation}\label{dotrHfix}
\dot{r}=\pb{r,H_{fix}}=\frac{A_{(p,q)}p_r}{\sqrt{\mK}} \ .
\end{equation}
To proceed further we use the fact that the Hamiltonian density
$\mH_{fix}$ is conserved and we denote its constant value  as $E$.
Then we can solve $\mH_{fix}=E$ for $p_r$ as
\begin{equation}
p_r=\sqrt{\frac{E^2-\frac{A_{(p,q)}L^2}{r^2}-T_{D1}^2 (m'^2+n'^2
A_{(p,q)})}{A_{(p,q)}}} \
\end{equation}
so that from (\ref{dotrHfix})  we obtain
\begin{equation}\label{dotrnonL}
\dot{r}^2=A_{(p,q)}- \frac{A^2_{(p,q)}}{E^2 }\left(\frac{L^2}{r^2}+
T_{D1}^2n'^2\right)-\frac{A_{(p,q)}T_{D1}^2}{E^2}m'^2 \ .
\end{equation}
As the check note  that the first two terms on the right side in
(\ref{dotrnonL})  coincide with the expression that governs the
dynamic of Dp-brane in NS5-brane background \cite{Kutasov:2004dj}
while the last one that is proportional to $m'^2$ corresponds to the
dynamics of the fundamental string in this background. For the next
purposes we also determine the equation of motion for $\phi$
\begin{equation}\label{dotphins}
\dot{\phi}=\pb{\phi,H_{fix}}=\frac{A_{(p,q)} L}{r^2 E} \ .
\end{equation}
\subsection{The Case $L=0$}
We firstly consider  the case of the vanishing angular momentum
$p_\theta=L=0$. Then the equation(\ref{dotphins}) implies that
$\phi$ is a constant while the (\ref{dotrnonL}) has the form
\begin{equation}\label{dotrfiveNS}
\dot{r}^2=A_{(p,q)}- \frac{A^2_{(p,q)}T_{D1}^2}{E^2}
n'^2-\frac{A_{(p,q)}T_{D1}^2}{E^2}m'^2 \ .
\end{equation}
Now we will analyze this expression in more details. First of all
the solution of this equation is restricted to the region where the
right side is non-negative. Since
 \begin{equation}
 A_{(p,q)}=\left(1+\frac{Q_{(p,q)}}{r^2}\right)^{-1} \ .
 \end{equation}
we obtain
\begin{equation}
\frac{Q_{(p,q)}}{ r^2}>\frac{T_{D1}^2n'^2}{E^2(1-\frac{T_{D1}^2}
{E^2}m'^2)}-1 \ .
\end{equation}
Note that for $m'=0$ this result agrees with the result derived in
\cite{Kutasov:2004dj}. We see that this condition is empty when
\begin{equation}
E^2>T_{D1}^2(n'^2+m'^2) \
\end{equation}
that has clear physical meaning. It corresponds to the situation
when the total energy is greater then the asymptotic tension of
$(m',n')$ string and given string can escape to infinity. Note that
for $E^2<T_{D1}^2(n'^2+m'^2)$ the $(m',n')$ string cannot escape the
attraction from five-brane.

We also determine the components of the stress energy tensor
(\ref{TNS}) for this configuration. Using (\ref{dotrfiveNS}) we
easily find
\begin{eqnarray}
T_{\tau\tau}&=&
\frac{T_{D1}\sqrt{m'^2+n'^2A_{(p,q)}}}{\sqrt{1-\frac{1}{A_{(p,q)}}
\dot{r}^2}}=E \ , \quad
T_{\tau\sigma}=0 \ , \nonumber \\
T_{\sigma\sigma}&=&-T_{D1} \sqrt{m'^2+n'^2 A_{(p,q)}}
\sqrt{1-\frac{1}{A_{(p,q)}}\dot{r}^2}=-\frac{T_{D1}^2}{E}
(m'^2+n'^2A_{(p,q)})\ . \nonumber \\
\end{eqnarray}
From $T_{\sigma\sigma}=\mP$ we see that  the contribution from
D1-brane to the pressure goes to zero when we approach the core of
the five-brane background   while the string like contribution is
constant. This is an analogue of the well known fact  that the
fundamental string can make the bound state with NS5-brane.

Let us now consider such an energy interval when the entire
trajectory  in the region when $Q_{(p,q)}\gg r^2$. Then the equation
for $\dot{r}$ has the form
\begin{eqnarray}\label{dotrnear}
\dot{r}^2=\frac{r^2}{Q_{(p,q)}}(1-\frac{T_{D1}^2m'^2}{E^2})
-\frac{r^4}{Q^2_{(p,q)}}\frac{T_{D1}^2}{E^2}n'^2 \nonumber \\
\end{eqnarray}
that has solution
\begin{equation}
r=\frac{1}{n'}\sqrt{Q_{(p,q)}\frac{E^2}{T_{D1}^2}-m'^2} \frac{1}{
\cosh \sqrt{\left(1-\frac{T_{D1}^2}{E^2}m'^2\right)
\frac{1}{Q_{(p,q)}}}t}  \ ,
\end{equation}
where we chosen the initial condition that for $t=0$
$(m',n')-$string is at the point of the maximal value corresponding
to $\dot{r}=0$. From the previous expression we see that this result
is valid in case of $m'=0$. On the other  hand the case  $n'=0$ has
to be analyzed separately in the equation (\ref{dotrnear}) and we
obtain the result
\begin{equation}
r=r_0e^{\pm\sqrt{\frac{1}{Q_{(p,q)}}(1-\frac{T_{D1}^2}{E^2}m'^2)}t}
\ ,
\end{equation}
where the $-$sign corresponds to $m'-$string moving towards to the
world-volume of five brane while $+$ corresponds to the situation
when $m'-$ string leaves it. Again, this result is the manifestation
 of the fact that the fundamental string can  form marginal bound
state with NS5-brane. However in our case this situation is not so
clear due to the fact that $m'$ is not an integer and depends on the
asymptotic values of $\Phi_0$ and $\chi_0$. On the other hand it is
clear that the equation of motion (\ref{dotrfiveNS}) possesses
constant solution $r=\mathrm{const}$ in case when $n'=0$ on
condition when
\begin{equation}\label{ETpq}
E^2=T_{D1}^2m'^2=T_{D1}^2(p^2+q^2) \ .
\end{equation}
This is rather puzzling result that shows the difficulty with the
background  solution (\ref{solRoy}). To see this in more details let
us imagine that we have  a configuration of the background NS5-brane
and probe fundamental string. Under $SL(2,Z)$ transformation these
two objects transform differently. Explicitly, since NS5-brane is
magnetically charged object with respect to NSNS-two form it
transforms in the same way as in (\ref{deftrB}). Then $(p,q)-$five
brane arises from $NS5-$brane through following $SL(2,Z)$
transformation
\begin{equation}
\left(\begin{array}{cc} \hat{Q}_{NS5} \\
\hat{Q}_{D5} \\ \end{array}\right)= \left(
\begin{array}{cc}
p & -c \\
q & a \\ \end{array} \right)
\left(\begin{array}{cc} 1 \\
0 \\ \end{array}\right)
\end{equation}
so that
\begin{equation}\label{pqtr}
\Lambda=\left(\begin{array}{cc} a & -q \\
c & p \\ \end{array}\right) \ .  \end{equation}
 On the other hand we know that
the fundamental string transforms under $SL(2,Z)$ transformation as
in (\ref{transbm}). Then we find that for $\Lambda$ given in
(\ref{pqtr}) we obtain $(a,c)-$string where $(ap+qc=1)$. Since
NS5-brane and fundamental string forms a marginal bound state the
previous arguments suggest that such a bound state exists also for
$(p,q)-$five brane and $(a,c)-$fundamental string. Then the
condition given  in  (\ref {transbm}) is not in agreement with this
claim. We mean that the resolution of this paradox can be found when
we construct the background $(p,q)-$five brane solution with the
help of $SL(2,Z)$ transformation rather than procedure used in
\cite{Lu:1998vh} that was based on the $SL(2,R)$ transformation.
This question is now under active investigation and we hope to
report our results soon.
\subsection{The Case $L\neq 0$}
Let us now consider the case  of non-zero angular momentum $L$.
Following \cite{Kutasov:2004dj} we rewrite the equation of motion
for (\ref{dotrnonL})
 into the form
\begin{equation}
\dot{r}^2+\frac{A^2_{(p,q)}}{E^2}\left(\frac{L^2}{r^2}+T_{D1}^2
n'^2\right)+A_{(p,q)}\frac{T_{D1}^2}{E^2}m'^2- A_{(p,q)}=0
\end{equation}
that can be interpreted as the equation of conserved   energy for a
particle with mass $m=2$ that moves in the effective potential
$V_{eff}(r)$
\begin{equation}
V_{eff}=\frac{A^2_{(p,q)}}{E^2}\left(\frac{L^2}{r^2}+T_{D1}^2
n'^2\right)+A_{(p,q)}\frac{T_{D1}^2}{E^2}m'^2- A_{(p,q)} \
\end{equation}
with zero energy.
 Now following \cite{Kutasov:2004dj}
 we will analyze the behavior of this
potential for different values of $r$. For small $r$ we obtain
\begin{equation}
V_{eff}=\frac{r^2}{Q_{(p,q)}}\left(\frac{L^2}{Q_{(p,q)}E^2}
+\frac{T_{D1}^2}{Q_{(p,q)}E^2}m'^2-1\right) \ .
\end{equation}
On the other hand for large $r$ we have
\begin{equation}
V_{eff}=\frac{T_{D1}^2}{E^2}n'^2-1 \ .
\end{equation}
Now we see that for $E<T_{D1}n'$ the potential $V_{eff}$ approaches
positive value for $r\rightarrow \infty$ and  since the particle has
zero  energy we find that it cannot escape to infinity. Further, in
order to have trajectories with non-zero $r$ we have to demand that
the potential approaches zero from below which implies
\begin{equation}\label{Lbound}
\frac{L^2}{Q_{(p,q)}}<E^2-\frac{T_{D1}^2m'^2}{Q_{(p,q)}} \ .
\end{equation}
In fact, if this condition were not satisfied that the only solution
would be $r=0$.

Let us now explicitly find the solution of the equation of motion in
the throat region when $A_{(p,q)}=\frac{r^2}{Q_{(p,q)}}$. Then the
equation (\ref{dotrnonL}) has the form
\begin{equation}
\dot{r}^2=\frac{r^2}{Q_{(p,q)}} \left(1-\frac{T_{D1}^2}{E^2}m'^2
-\frac{L^2}{Q_{(p,q)}E^2}\right)-\frac{r^4}{Q^2_{(p,q)}E^2}T_{D1}^2
n'^2 \
\end{equation}
that has the solution
\begin{equation}
r=\frac{Q_{(p,q)}E}{T_{D1}} \sqrt{1-m'^2\frac{T_{D1}^2}{E^2}
-\frac{L^2}{Q_{(p,q)}E^2}}\frac{1}{\cosh\sqrt{1-\frac{T_{D1}^2}{E^2}m'^2
-\frac{L^2}{Q_{(p,q)}E^2}}t} \ .
\end{equation}
We see that the non-zero angular momentum slows down the decrease of
$r$. Further,the equation of motion for $\phi$ implies
\begin{equation}
\phi=\frac{L}{EQ_{(p,q)}}t \ .
\end{equation}
In other words, previous solution describes $(m',n')-$string that
moves towards to the world-volume of background five brane
 which however also circles around them.

As the next example we  consider the situation  when $n'=0$. In this
case we find the potential in the form
\begin{equation}
V_{eff}=\frac{A^2_{(p,q)}}{E^2}\frac{L^2}{r^2}
+A_{(p,q)}\frac{T_{D1}^2}{E^2}m'^2- A_{(p,q)}
\end{equation}
that in the throat region simplifies as
\begin{equation}
V_{eff}=A_{(p,q)}\left(\frac{L^2}{Q_{(p,q)}E^2}+\frac{T_{D1}^2}{E^2}m'^2-1\right)
\end{equation}
and it vanishes identically when
\begin{equation}
E^2=\frac{L^2}{Q_{(p,q)}}+T_{D1}^2 m'^2 \ .
\end{equation}
In other words it is possible to find string that rotates around
five brane for any values of $r$.

The situation is different  when $E>T_{D1}n'$ which means that the
potential is negative for $r\rightarrow \infty$. Further, if we
again have (\ref{Lbound}) we obtain that we approach the point $r=0$
from below and hence there is no potential barrier. In this case we
have a possibility of the particle that starts  at $r=0$ for
$t=-\infty$ and it escapes to infinity in  time reverse process. On
the other hand the situation is different when the bound
(\ref{Lbound}) is not satisfied. Let us imagine that we have
$(m,n)-$string initially at large distance from $(p,q)-$five brane.
The probe moves towards to the $(p,q)-$five brane until it reaches
the point when the effective potential vanishes that is at
\begin{equation}
r_{min}^2=\frac{L^2-E^2 Q_{(p,q)}-T^2_{D1}m'^2 Q_{(p,q)}}
{E^2-T_{D1}^2 m'^2-T^2_{D1}n'^2} \ .
\end{equation}
Following \cite{Kutasov:2004dj} we can interpret this  process as a
scattering of $(m,n)-$string from the collection o $(p,q)-$five
branes. Since the analysis is completely the same as in
\cite{Kutasov:2004dj} we will not repeat it here.

\section{$(m,n)$-String in $(p,q)$-String Background}\label{fourth}
In this section we consider dynamics of $(m,n)-$string in the
macroscopic $(p,q)-$ string background \cite{Schwarz:1995dk}
\begin{eqnarray}
ds^2_E&=&H_{pq}^{-3/4}[-dt^2+dy^2]+H_{pq}^{1/4} dx_m dx^m \ ,
\quad
H_{pq}=1+\frac{(2\pi)^6 \alpha'^3\triangle_{(p,q)}^{1/2}}{ r^6
\Omega_7}\equiv 1+\frac{\alpha}{r^6} \ , \nonumber \\
\bB&=&(\Lambda^{-1}_{p,q})^T\left(\begin{array}{cc}( H_{pq}^{-1}-1) \\
0 \\ \end{array}\right) \ , \quad
\Lambda=\triangle^{-1/2}_{(p,q)}\left(\begin{array}{cc} p &
-e^{-\Phi_0}q+\chi_0 e^{\Phi_0}(p-q\chi_0) \\
q & e^{\Phi_0}(p-q\chi_0) \\
\end{array}\right) \ ,  \nonumber \\
\mM&=&\Lambda \left(\begin{array}{cc} H_{pq}^{1/2} & 0 \\
0 & H^{-1/2}_{pq} \\ \end{array}\right) \Lambda^T \ , \quad
\triangle_{(p,q)}=e^{-\Phi_0}q^2+(p-q\chi_0)^2e^{\Phi_0} \ . \nonumber \\
\end{eqnarray}
The Hamiltonian density for the time dependent world-sheet modes has the form
\begin{eqnarray}\label{Hamfixstring}
\mH_{fix}=\sqrt{\Pi_m \Pi^m H^{-1}_{pq}+H^{-2}_{pq}T_{D1}^2
(m'^2+n'^2H_{pq})}-T_{D1}m'(H_{pq}^{-1}-1) \ , \nonumber \\
\end{eqnarray}
where $\bm'$ is equal to
\begin{equation}
\bm'=\left(\begin{array}{cc}
m'\\
n' \\ \end{array}\right)=\triangle^{-1/2}_{(p,q)}
\left(\begin{array}{cc}
e^{\Phi_0}(p-q \chi_0)(m-\chi_0 n)+e^{-\Phi_0} qn \\
-qm+pn \\ \end{array}\right) \ .
\end{equation}
Clearly for $m=p$ and $n=q$ we obtain
\begin{equation}\label{bmprobestring}
\bm'=\triangle^{-1/2}_{(p,q)}\left(\begin{array}{cc}
e^{\Phi_0}(p-q\chi_0)^2+e^{-\Phi_0} qn \\
0 \\ \end{array}\right)=\left(\begin{array}{cc}
\triangle^{1/2}_{(p,q)} \\
0 \\ \end{array}\right)
\end{equation}
with following physical interpretation. As we know $(p,q)-$string
solution was derived through $SL(2,R)$ transformation from the
fundamental string background. Then clearly a fundamental string
 in a macroscopic string background maps to the same object under
$SL(2,Z)$ transformation. On the other hand from
(\ref{bmprobestring}) we see that this is not exactly true since the
probe string  does not carry  integer charge. We again leave the
resolution of this paradox for future research.

It is also useful to find components of the stress energy tensor for
the $(m,n)-$string in static gauge. As in the previous section we
temporary replace fixed two dimensional metric $\eta_{\alpha\beta}$
with two dimensional metric $\gamma_{\alpha\beta}$ and write the
gauge fixed action in the form
\begin{equation}
S_{fixed}=-T_{D1}\int d\tau d\sigma\left(\sqrt{\bm^T\mM^{-1} \bm
}H_{pq}^{-3/4}\sqrt{-\det \bA} - \sqrt{-\gamma}
(\Lambda^{-1}_{pq}\bm)^T\left(\begin{array}{cc}
\frac{1}{H_{pq}}-1 \\
0 \\ \end{array}\right)\right) \ ,
\end{equation}
where
\begin{equation}
\bA_{\alpha\beta}=\gamma_{\alpha\beta}+H_{pq}\delta_{mn}
\partial_\alpha x^m\partial_\beta x^n \ .
\end{equation}
Then the components of two dimensional stress energy tensor have the
form
\begin{eqnarray}
T_{\alpha\beta}&=&-\frac{2}{\sqrt{-\gamma}}\frac{\delta
S_{fixed}}{\delta \gamma^{\alpha\beta}}=\nonumber
\\
&=&-\frac{T_{D1}}{\sqrt{-
\gamma}}\gamma_{\alpha\gamma}(\bA^{-1})^{\gamma\delta}\gamma_{\delta\beta}
\sqrt{-\det\bA}\sqrt{\bm^T\mM^{-1} \bm }H_{pq}^{-3/4}+
\gamma_{\alpha\beta}(\Lambda^{-1}_{pq}\bm)^T\left(\begin{array}{cc}
\frac{1}{H_{pq}}-1 \\
0 \\ \end{array}\right) \ . \nonumber \\
\end{eqnarray}
Finally we return back to the flat metric
$\gamma_{\alpha\beta}\rightarrow \eta_{\alpha\beta}$ and consider
pure time-dependent ansatz. Then we obtain
\begin{eqnarray}\label{setFS}
T_{\tau\tau}
&=& \frac{T_{D1}\sqrt{m'^2+n'^2H_{pq}}}{H_{pq} \sqrt{
1-H_{pq}(\dot{r}^2+r^2\dot{\phi}^2)}}+m'T_{D1}(1-\frac{1}{H_{pq}})
 \ , \quad
T_{\tau\sigma}=0 \ , \nonumber \\
T_{\sigma\sigma}
&=& -T_{D1}\sqrt{m'^2+n'^2H_{pq}} \frac{1}{H_{pq}}
\sqrt{1-H_{pq}(\dot{r}^2+r^2\dot{\phi}^2)}-m'T_{D1}\left(1-\frac{1}{H_{pq}}\right)
 \ , \nonumber \\
\end{eqnarray}
where we also introduced  spherical coordinates in the transverse
$\mathbf{R}^8$ space and considered the dynamics of the probe in two
dimensional plane with radial variable $r$ and angular $\phi$. As a
result the Hamiltonian density (\ref{Hamfixstring}) simplifies
considerably
\begin{equation}
\mH_{fix}=\frac{1}{H_{pq}} \left(\sqrt{
H_{pq}(p_r^2+\frac{1}{r^2}p_\phi^2+T_{D1}^2 n'^ 2)+T_{D1}^2m'^
2}+T_{D1}m'(H_{pq}-1)\right) \ .
\end{equation}
   Note also that
the equation of motion for $\phi$ has the form
\begin{equation}
\dot{\phi}=\pb{\phi,H_{fix}}=
\frac{L}{r^2(H_{pq}E-T_{D1}m'(H_{pq}-1))} \ ,
\end{equation}
where we used the fact that  $p_\phi=L$ and $\mH_{fix}=E$ are
conserved. With the help of these results we obtain
following components of the stress energy
tensor (\ref{setFS})
\begin{eqnarray}
T_{\tau\tau}&=&E \ , \nonumber \\
T_{\sigma\sigma}&=&\mP =-\frac{T_{D1}^2(m'^2+n'^2H_{pq})}{
H_{pq}(E-T_{D1}m')+T_{D1}m'}-T_{D1}m'\left(1-\frac{1}{H_{pq}}\right)
\ ,
\end{eqnarray}
where $\mP$ is the pressure on the world-volume of $(m,n)$-string.

%
Let us now proceed to the analysis of dynamics of this probe string.
From $H_{fix}$ we derive the equation of motion
\begin{equation}\label{dotrfs}
\dot{r}=\pb{r,H_{fix}}=\frac{p_r}{\sqrt{
H_{pq1}(p_r^2+\frac{1}{r^2}L^2+T_{D1}^2 n'^ 2)+T_{D1}^2m'^2}} \ .
\end{equation}
On the other hand from the fact that the energy density is conserved
$E$ we can express $p_r$ as
\begin{eqnarray}
p_r^2
=\frac{1}{H_{pq}}\left(
(H_{pq}(E-T_{D1}m')+m'T_{D1})^2-T_{D1}^2m'^2\right)-\frac{L^2}{r^2}
-T_{D1}^2 n'^2  \nonumber \\
\end{eqnarray}
so that (\ref{dotrfs}) has the form
\begin{eqnarray}
\dot{r}^2
=\frac{1}{H_{pq}}\left(1-\frac{(T_{D1}^2n'^2+\frac{L^2}{r^2})H_{pq}+T_{D1}^2
m'^2}{(H_{pq}(E-T_{D1}m')+T_{D1}m')^2}\right) \ .  \nonumber \\
\end{eqnarray}
We can again rewrite this equation into the more suggestive form
\begin{equation}
 \dot{r}^2+V_{eff}=0 \ ,
\end{equation}
where
\begin{equation}\label{Veffstring}
V_{eff}=\frac{1}{H_{pq}}\left(\frac{(T_{D1}^2n'^2+\frac{L^2}{r^2})H_{pq}+T_{D1}^2
m'^2}{(H_{pq}(E-T_{D1}m')+T_{D1}m')^2}-1\right) \ .
\end{equation}
We see that the equation above corresponds to the massive particle
with mass $m=2$ moving in the potential $V_{eff}$ with zero energy
so that many interesting information about the particle's trajectory
follow from the  properties of given potential.
 As in previous section
we start with the case of the zero angular momentum
\subsection{The Case $L=0$}
We see that for  $r\rightarrow \infty$ we have $H_{pq}\rightarrow 1$
and hence
\begin{equation}\label{Veffinfty}
V_{eff}\rightarrow \frac{T_{D1}^2 m'^2+T_{D1}^2 n'^2}{E^2}-1 \
\end{equation}
while for small $r$  we obtain
\begin{equation}\label{Veffnearstring}
V_{eff}=\frac{r^6}{\alpha}\left(\frac{n'^2T_{D1}^2r^6}
{\alpha (E-T_{D1}m')^2}-1\right)
\end{equation}
so that  $V_{eff}$ approaches the point $r=0$  from below. As a
result we have two qualitative different  behaviors of probe
$(m',n')$ string in this background. It follows from
(\ref{Veffinfty}) that for $E^2<T_{D1}^2(m'^2+n'^2)$ $V_{eff}$
approaches positive constant for large $r$. Then the probe string
cannot escape to infinity and moves in the bounded region around
$(p,q)$-string background. On the other hand for $E^2>T_{D1}^2(m'^2+
n^2)$ the potential is negative for all values and hence probe
string can moves to infinity. Let us firstly consider the case when
$n'=0$. This case corresponds to the situation of the motion of
fundamental string in the background of collection of the
fundamental strings. We can expect that it is possible to form a
marginal bound state of $N+m'$ fundamental strings. In fact, for
$E-T_{D1}m'=\epsilon\ll 1$ we find that the effective potential has
the form
\begin{equation}
V_{eff}=-2\frac{\epsilon}{T_{D1}m'}
\end{equation}
and we see that it  is flat. As a result we find time dependence
$r\sim \pm\epsilon t$ which means very slowly movement of the probe
string. This is a confirmation of the claim that the probe string
can form marginal bound state with $N$ fundamental strings. Note
that this approximation is valid on condition when
\begin{equation}
\frac{H_{pq}\epsilon}{T_{D1}m'}\ll 1
\end{equation}
that implies $r^6\gg \frac{\alpha \epsilon}{T_{D1} m'}$ that can be
obeyed  in the whole region in the limit $\epsilon\rightarrow 0$.
Finally note also that the pressure is equal to
\begin{equation}
\mP=-2T_{D1}m'+H_{pq}\epsilon +\frac{T_{D1}}{H_{pq}}m' \
\end{equation}
that has following physical explanation. Consider the initial
configuration when the  $m'-$string is sitting  at infinity when
$H_{pq}=1$ and consequently $\mP=-T_{D1}m'+\epsilon=E-2T_{D1}m'$.
The string moves slowly to the horizon when $H_{pq}\rightarrow
\infty$ and hence the pressure approaches the value $\mP\rightarrow
-m'T_{D1}$ at the horizon in the limit $\epsilon\rightarrow 0$.

To see this more clearly let us consider the case of the near
horizon limit when $H_{pq}\epsilon\gg T_{D1}m'$ when $\epsilon$ is
small but finite. In this case we find that the leading order
behavior of the effective potential is
\begin{equation}
V_{eff}=-\frac{r^6}{\alpha}
\end{equation}
and hence we have the differential equation
\begin{equation}
\frac{dr}{dt}=\pm\frac{r^3}{\sqrt{\alpha}} \ .
\end{equation}
The $+$ sign corresponds to the string moving from the horizon when
the approximation we use quickly breaks. The sign $-$ corresponds to
the string moving to the horizon and for this possibility we find
the solution
\begin{equation}
r=\frac{r_0}{\sqrt{1+\frac{2r_0^2}{\sqrt{\alpha}}t}} \ , \quad
r_0^6\ll \alpha  \ .
\end{equation}
We see that the probe string approaches horizon at asymptotic time
$t\rightarrow \infty$. Observe that this behavior does not depend on
the value of the energy of the string probe.
\subsection{The Case  $L\neq 0$}
Now we would like to analyze the behavior of the potential in case
$n'=0$ and $L\neq 0$ and in the limit  $E-T_{D1}m'=\epsilon\ll 1$.
In this case we find the effective potential in the form
\begin{eqnarray}\label{Veffapr}
V_{eff}
&=&-2\frac{\epsilon}{T_{D1}m'}+\frac{L^2}{r^2T_{D1}^2
m'^2}(1-\frac{2\epsilon}{T_{D1}m'})-\frac{2\alpha}{r^8}
\frac{L^2\epsilon}{T_{D1}^3m'^3} \ . \nonumber \\
\end{eqnarray}
From this form of the effective potential we can deduce an existence
of the potential barrier since there are two points where $V_{eff}$
vanishes. We find these points as follows. We presume that the first
root corresponds to the root when we neglect term proportional to
$r^{-8}$. Then we solve the quadratic equation with the solution
\begin{equation}
r_+=\frac{L}{\sqrt{2T_{D1}m'\epsilon}} \ .
\end{equation}
We see that it has very large value that justifies our assumption.
The second root corresponds to the situation when we neglect the
constant term in (\ref{Veffapr}) and we obtain
\begin{equation}
r_-=\left(\frac{2\alpha\epsilon}{T_{D1}m'}\right)^{1/6}
\end{equation}
that is much smaller than $r_+$ again with agreement with our
assumptions.  The physical picture is following. If we have probe
$m'-$string with $E-T_{D1}m'\ll 1$ far away from the background
$(p,q)-$string that it moves towards it however it cannot cross the
horizon. Rather it approaches to the distance given by $r_+$ and
than it is deflected. On the other hand $m'-$ string that is
initially in the region below $r_-$ will spirally moves towards to
the horizon. This situation is similar as in case of $(m,1)$-string
studied in \cite{Bak:2004tp} and we will not repeat it here.

\vskip .5in \noindent {\bf Acknowledgement:}
\\
This work  was supported by the Grant Agency of the Czech Republic
under the grant P201/12/G028.


\begin{thebibliography}{20}

\bibitem{Duff:1994an}
  M.~J.~Duff, R.~R.~Khuri and J.~X.~Lu,
\emph{``String solitons,''}
  Phys.\ Rept.\  {\bf 259} (1995) 213
  doi:10.1016/0370-1573(95)00002-X
  [hep-th/9412184].





\bibitem{Smith:2002wn}
  D.~J.~Smith,
\emph{``Intersecting brane solutions in string and M theory,''}
  Class.\ Quant.\ Grav.\  {\bf 20} (2003) R233
  doi:10.1088/0264-9381/20/9/203
  [hep-th/0210157].

\bibitem{Stelle:1996tz}
  K.~S.~Stelle,
\emph{``Lectures on supergravity p-branes,''}
  In *Trieste 1996, High energy physics and cosmology* 287-339
  [hep-th/9701088].


\bibitem{Ortin:2004ms}
  T.~Ortin,
\emph{``Gravity and strings,''}

\bibitem{Tseytlin:1996bh}
  A.~A.~Tseytlin,
\emph{``Harmonic superpositions of M-branes,''}
  Nucl.\ Phys.\ B {\bf 475} (1996) 149
  doi:10.1016/0550-3213(96)00328-8
  [hep-th/9604035].

\bibitem{Obers:1998fb}
  N.~A.~Obers and B.~Pioline,
\emph{``U duality and M theory,''}
  Phys.\ Rept.\  {\bf 318} (1999) 113
  doi:10.1016/S0370-1573(99)00004-6
  [hep-th/9809039].


\bibitem{Schwarz:1995dk}
  J.~H.~Schwarz,
\emph{``An SL(2,Z) multiplet of type IIB superstrings,''}
  Phys.\ Lett.\ B {\bf 360} (1995) 13
   [Phys.\ Lett.\ B {\bf 364} (1995) 252]
  doi:10.1016/0370-2693(95)01138-G
  [hep-th/9508143].




\bibitem{Lu:1998vh}
  J.~X.~Lu and S.~Roy,
\emph{``An SL(2,Z) multiplet
 of type IIB super five-branes,''}
  Phys.\ Lett.\ B {\bf 428} (1998) 289
  doi:10.1016/S0370-2693(98)00435-3
  [hep-th/9802080].



\bibitem{Tseytlin:1996it}
  A.~A.~Tseytlin,
\emph{``Selfduality of Born-Infeld action and Dirichlet three-brane
of type IIB superstring theory,''}
  Nucl.\ Phys.\ B {\bf 469} (1996) 51
  doi:10.1016/0550-3213(96)00173-3
  [hep-th/9602064].




\bibitem{Lozano:1997cy}
  Y.~Lozano,
\emph{``D-brane dualities as canonical transformations,''}
  Phys.\ Lett.\ B {\bf 399} (1997) 233
  doi:10.1016/S0370-2693(97)00292-X
  [hep-th/9701186].



\bibitem{Cederwall:1997ts}
  M.~Cederwall and P.~K.~Townsend,
\emph{``The Manifestly Sl(2,Z) covariant superstring,''}
  JHEP {\bf 9709} (1997) 003
  doi:10.1088/1126-6708/1997/09/003
  [hep-th/9709002].

\bibitem{Townsend:1997kr}
  P.~K.~Townsend,
\emph{``Membrane tension and manifest IIB S duality,''}
  Phys.\ Lett.\ B {\bf 409} (1997) 131
  doi:10.1016/S0370-2693(97)00862-9
  [hep-th/9705160].


\bibitem{Bergshoeff:2006gs}
  E.~A.~Bergshoeff, M.~de Roo, S.~F.~Kerstan, T.~Ortin and F.~Riccioni,
\emph{``SL(2,R)-invariant IIB Brane Actions,''}
  JHEP {\bf 0702} (2007) 007
  doi:10.1088/1126-6708/2007/02/007
  [hep-th/0611036].

\bibitem{Kutasov:2004dj}
  D.~Kutasov,
\emph{``D-brane dynamics near NS5-branes,''}
  hep-th/0405058.

\bibitem{Bak:2004tp}
  D.~Bak, S.~J.~Rey and H.~U.~Yee,
\emph{``Exactly soluble dynamics of
(p,q) string near macroscopic fundamental strings,''}
  JHEP {\bf 0412} (2004) 008
  doi:10.1088/1126-6708/2004/12/008
  [hep-th/0411099].

\bibitem{Kluson:2014uaa}
  J.~Kluson,
\emph{``Integrability of D1-brane on Group Manifold,''}
  JHEP {\bf 1409} (2014) 159
  doi:10.1007/JHEP09(2014)159
  [arXiv:1407.7665 [hep-th]].

\bibitem{Bueno:2014zja}
  P.~Bueno, T.~Ortin and C.~S.~Shahbazi,
\emph{``Non-extremal branes,''}
  Phys.\ Lett.\ B {\bf 743} (2015) 301
  doi:10.1016/j.physletb.2015.02.070
  [arXiv:1412.5547 [hep-th]].

\end{thebibliography}
\end{document}